\documentclass[onecolumn, a4paper,journal,10pt]{IEEEtran}
\usepackage{amsfonts}
\usepackage{graphicx, epsfig,amsmath,amssymb,latexsym,graphics,float}
\usepackage{setspace,subfig}
\usepackage{citesort}
\usepackage{float}

\include{IEEEtran.bib}

\begin{document}

\title{Posterior Exploration based Sequential Monte Carlo for Global Optimization}
\author{Bin~Liu
\thanks{B. Liu is with Nanjing University of Posts and Telecommunications, Nanjing,
Jiangsu, 210023 China e-mail: bins@ieee.org.}
\thanks{Manuscript received Nov. 1, 2015; revised July 10, 2016.}}
\maketitle

\begin{abstract}
We propose a global optimization algorithm based on the Sequential Monte Carlo (SMC) sampling framework. In this framework, the objective function is normalized to be a probabilistic density function (pdf), based on which a sequence of annealed target pdfs is designed to asymptotically converge on the set of global optima. A sequential importance sampling (SIS) procedure is performed to simulate the resulting targets, and the maxima of the objective function is assessed from the yielded samples. The disturbing issue lies in the design of the importance sampling (IS) pdf, which crucially influences the IS efficiency. We propose an approach to design the IS pdf online by embedding a posterior exploration (PE) procedure into each iteration of the SMC framework. The PE procedure can also explore the important regions of the solution space supported by the target pdf. A byproduct of the PE procedure is an adaptive mechanism to design the annealing temperature schedule online. We compare the performance of the proposed algorithm with those of several existing related alternatives by applying them to over a dozen standard benchmark functions. The result demonstrates the appealing properties of our algorithm.
\end{abstract}


\section{Introduction}\label{sec:intro}
\IEEEPARstart{S}{imulation} based algorithms consist of a branch of population-based approaches to solve optimization problems. Different from the other branches of population-based methods such as particle swarm optimization \cite{kennedy2010particle,poli2007particle}, genetic algorithm \cite{whitley1994genetic} and ant colony optimization \cite{dorigo2010ant,dorigo2006ant}, the basic idea of simulation based algorithms is to simulate a particular probabilistic density function (pdf) over the promising areas of the global optimum. In particular, if the objective function $f$ is non-negative and, consequently, we search for its maximum, we can sample from a pdf $\pi(\cdot)$, resulting from the normalization of $f$, since the mode of the resulting distribution will capture the maximum point we are searching for.

The degree of difficulty in using simulation based methods to find the global optimum largely depends on the structure of the normalized objective function $\pi(\cdot)$. If $\pi(\cdot)$ is a one dimensional unimodal distribution, it only needs to run dozens of steps of a benchmark Metropolis-Hastings sampling to get a satisfactory answer. If $\pi(\cdot)$ is a multivariate multimodal distribution, it could become challenging to develop a suitable simulation algorithm to yield a satisfactory answer fast.

There are mainly two directions to resolve the above challenge. The first direction is to introduce a set of smoothly transitional target distributions, which is expected to asymptotically converge on the set of global optima, and then simulate the target distributions sequentially. The motivation is to track closely the converging sequence of distributions. The celebrated simulated annealing (SA) algorithm just falls into the above framework. In particular, the standard SA applies the Boltzmann distributions, which is featured by a parameter termed annealing temperature, to serve as the transitional target distributions \cite{kirkpatrick1983optimization,ingber1993simulated}. At each iteration, SA essentially simulates a Markov chain whose stationary distribution is the Boltzmann distribution of the current temperature, and the current state becomes the initial state for a new chain at the next iteration. Recently a Sequential Monte Carlo-Simulated Annealing (SMC-SA) algorithm has been proposed, in which the Markov Chain Monte Carlo (MCMC) procedure embedded in the standard SA algorithm is replaced by an importance sampling (IS) procedure \cite{zhou2013sequential}. Different from MCMC, the IS procedure allows for parallel implementations and easy assessment of the Monte Carlo error \cite{oh1992adaptive,cappe2008adaptive}. On the other hand, the main shortcoming of IS approaches is an acute sensitivity to the choice of the IS pdf combined with the fact that it is impossible to come up with a universally efficient IS density \cite{oh1992adaptive,cappe2008adaptive}.
Further, in all SA based methods, the temperature has to decrease slowly enough such that the target distribution does not vary too much from iteration to iteration, which ensures the overall convergence \cite{neal2001annealed,del2006sequential}. However a slowly decreasing temperature schedule indicates more iterations, leading to more computing burdens. In practice the temperature schedule is often specified beforehand by human experts, while it is desirable if an adaptive mechanism can be developed to yield an appropriate temperature schedule online.

The second direction to resolve the above challenge resulted from a complex $\pi(\cdot)$ is to design a model that explicitly defines a probability distribution $q$ over candidate solutions. The estimation of distribution algorithms (EDAs) \cite{hauschild2011introduction}, evolutionary strategy (ES) \cite{auger2012tutorial,krause2015more,beyer2012design}, cross entropy (CE) methods \cite{rubinstein2013cross,de2005tutorial} and the particle filter optimization (PFO) algorithm \cite{zhou2014particle,chen2013population,liu2016particle} all belong to this branch, and thus we call them model based approaches in what follows. In these approaches, a random data set is first generated and then a model is created based on this random data set. The model is then sampled to generate new candidate solutions. At each iteration,  the model is updated in order to increase the model quality, ensuring that it will generate better candidate solutions in the following iteration. In EDAs, probabilistic models such as probability vectors and Bayesian networks are built from samples of high quality solutions \cite{hauschild2011introduction}. In ES, new candidate solutions are often generated from a single normal distribution centered around the best-so-far candidate solution or from a mixture of normal distributions centered around a population of high-quality solutions that has been found \cite{auger2012tutorial}.
For both the EDAs and ES, there is no explicit definition of the model quality. Different from them, CE methods define the quality of a model in terms of the Kullback-Leibler (KL) distance. It is a non-symmetric measure of the difference between two probability distributions \cite{rubinstein2013cross,de2005tutorial}. Specifically, CE methods try to find a pdf that is closer to $\pi(\cdot)$ as much as possible and the closeness is measured by the KL distance. The PFO algorithm is proven to be equivalent with CE methods, see details in \cite{zhou2014particle}. The efficiency of such model based approaches largely depends on the choice of the model structure and the model parameter updating mechanism. A labor intensive tuning process is usually required before running such algorithms. Further, they are often trapped into local optimum in the searching process.

Motivated by the observation that both the above two directions have their respective limitations, in this paper, we develop a novel simulation based algorithm, termed posterior exploration based SMC (PE-SMC). This algorithm improves state-of-the-art SMC-SA algorithm \cite{zhou2013sequential} by fusing a posterior exploration procedure into the framework of SMC. This PE procedure has three functions. First, it can update the IS density model adaptively at each iteration of SMC. Second it can explore the promising regions of the candidate solutions in both macroscopic and microscopic scales. Lastly, the output of the PE procedure can be used for selecting the annealing temperature of the following iteration.

The remainder of the paper is organized as follows. In Section II, we give a brief overview on the SMC-SA optimization algorithm \cite{zhou2013sequential}. In Section III, we present the proposed PE-SMC algorithm. In Section IV, we give a convergence analysis for the PE-SMC algorithm. In Section V, we evaluate the performance of our algorithm in solving a number of benchmark optimization problems. Finally, we conclude the paper in Section VI.
\section{Revisiting SMC-SA}
In this section, we present a brief overview on the SMC-SA algorithm \cite{zhou2013sequential}. The purpose is to define the notions and introduce the necessary background information for describing the proposed PE-SMC algorithm in Section III.
In this paper, we are concerned with the following maximization problem
\begin{equation}
\underset{x\in\chi}{\max}f(x)
\end{equation}
where $\chi$ denotes the nonempty solution space defined in $\mathbb{R}^n$, and $f$: $\chi\rightarrow\mathbb{R}$ is a continuous real-valued function. The basic assumption here is that $f$ is bounded on $\chi$, which means $\exists f_l>-\infty, f_u<\infty$ such that $f_l\leq f(x)\leq f_u$, $\forall x\in\chi$. We denote the maximal function value as $f^{\ast}$, i.e., there exists an $x^{\ast}$ such that $f(x)\leq f^{\ast}\triangleq f(x^{\ast})$, $\forall x\in\chi$.

A sequence of target pdfs, $\pi_1(f), \pi_2(f), \ldots,$ is built as follows
\begin{equation}
\pi_k(x)\triangleq \exp\left(\frac{f(x)}{T_k}\right), k=1,2,\ldots,
\end{equation}
where $T_k$ denotes the cooling temperature at the $k$th iteration satisfying $T_k<T_{k-1}$ if $k>1$. A working flow of the algorithm is summarized as follows.

Algorithm 1: SMC-SA
\begin{itemize}
\item Initialization (i.e., at iteration 0): Generate $x_0^i\overset{\mbox{iid}}{\thicksim}\mbox{Unif}(\chi)$, $i=1,2,\ldots,N_0$. $N_0$ and $N_k$ in what follows denote the sample size and belong to the input of the algorithm. Set $k=1$.
\item At iteration $k$:
\begin{itemize}
\item Importance weight updating: calculate normalized importance weights according to $\omega_k^i\propto \pi_1(x_0^i)$ if  $k=1$, and $\omega_k^i\propto \frac{\pi_k(x_{k-1}^i)}{\pi_{k-1}(x_{k-1}^i)}$ if $k>1$.
\item Resampling: draw i.i.d. samples $\{\tilde{x}_k^i\}_{i=1}^{N_k}$ from a distribution $\tilde{\pi}_k(x)\triangleq\sum_{i=1}^{N_{k-1}}\omega_k^i \delta(x-x_{k-1}^i)$, where $\delta$ denotes the Dirac delta function. 
\item One step of Metropolis sampling:
\begin{itemize}
\item  Choose a symmetric proposal distribution with density $g_k(\cdot|x)$, such as a normal distribution with mean $x$.
\item Generate $y_k^i\sim g_k(y|\tilde{x}_k^i)$, $i=1,\ldots,N_k$.
\item Calculate acceptance probability
\begin{displaymath}
\rho_k^i=\min\left\{\frac{\pi_k(y_k^i)}{\pi_k(\tilde{x}_k^i)},1\right\}.
\end{displaymath}
\item Accept
\begin{displaymath}
x_k^i=\left\{\begin{array}{ll}
y_k^i, \mbox{with probability}\quad\rho_k^i;\\
\tilde{x}_k^i, \mbox{with probability}\quad 1-\rho_k^i. \end{array} \right.
\end{displaymath}
\end{itemize}
\item Stopping: if a stopping criterion is satisfied, return $\max_i f(x_k^i)$;
otherwise, set $k=k+1$ and continue.
\end{itemize}
\end{itemize}

As is shown above, the SMC-SA algorithm tracks the converging sequence of Boltzmann
distributions using a population of samples via SMC method, such that the empirical
distributions yielded by SMC-SA also converge weakly to the uniform distribution concentrated on the set of global optima as the temperature decreases to zero.

A convergence analysis for the SMC-SA algorithm is given in \cite{zhou2013sequential}, which indicates that an appropriate choice of the sample size sequence $\{N_k\}$ and the temperature cooling schedule $\{T_k\}$ is required. There is a fundamental trade-off between the temperature change rate and the sample size. Conceptual guidance is provided in \cite{zhou2013sequential} for selecting the cooling schedule and the sample size, while it still lacks a specific working method for use in practice.

In addition, it is shown that the SMC-SA algorithm only depends on one step of Metropolis sampling procedure to generate new candidate solutions, while this operation only moves each candidate solution to its local neighborhood in the solution space $\chi$ with a probability.
Such a local searching mechanism is not good enough for finding the global optima, especially when the optimal solution lies far from the neighborhood of the solutions found so far; because the search may stop due to the impossibility of improving the solution, even if the best solution found so far is not optimal. Lastly the Boltzmann distributions may lead to numerical problems due to their exponential form, as pointed out in \cite{zhou2013sequential}.
\section{The proposed PE-SMC algorithm for global optimization}
In this section, we present the PE-SMC algorithm in detail.
The objective function $f(x)$ is assumed to be positive for any $x$.
The series of target pdfs is defined to be
\begin{equation}\label{anneal_fun}
\pi_k(x)\varpropto f(x)^{\lambda_k}, k=1,2,\ldots,
\end{equation}
where $k$ denotes the index of the target pdf, $\lambda\in\mathbb{R}$ the annealing temperature variable which satisfies $0<\lambda_{k-1}<\lambda_k$.
A working flow of our algorithm is summarized as follows.

Algorithm 2: PE-SMC
\begin{itemize}
\item Initialization (iteration 0): initialize the IS density function $q$ with parameter value $\psi_0$. Denote the resulting pdf by $q(\cdot|\psi_0)$. 
 Note that we select here a mixture of Student's $t$ pdfs to model $q$, see details in Sec.\ref{sec:Init}. Initialize the sample size $N$ and the cooling temperature $\lambda_1$. Set $k=1$.
\item At iteration $k$:
\begin{itemize}
\item Draw i.i.d. random samples $x_k^i$, $i=1,2,\ldots,N$ from $q(\cdot|\psi_{k-1})$.
\item Perform the PE procedure and output the revised parameter of $q$, denoted by $\psi_k$. See details in Sec.\ref{sec:PE}.
\item Select the cooling temperature for the following iteration, namely $\lambda_{k+1}$. See details in Sec.\ref{sec:temp_adapt}.
\item Stopping: if a stopping criterion is satisfied, return the best solution found so far, 
otherwise, set $k=k+1$ and continue.
\end{itemize}
\end{itemize}

In what follows, we introduce each module of the algorithm in detail.
\subsection{Initialization}\label{sec:Init}
We design the IS density $q$ to be a mixture of Student's $t$ pdfs. The use of a mixture type model makes $q$ easy to account for possible multimodal structures of the target pdfs. The Student's $t$ pdf is chosen as the mixing components due to its heavy-tail property, which is beneficial for the algorithm to explore the solution space.

Following \cite{liu2014adaptive}, suppose that $Y$ is a $d$ dimensional random variable that follows the multivariate Student's $t$ distribution, denoted by $\mathcal{S}(\cdot|\mu,\Sigma,v)$, where $\mu$ denotes the mean, $\Sigma$ a positive definite inner product matrix and $v\in(0,\infty]$ is the degrees of freedom. Then the density function of $Y$ is:
\begin{equation}\label{def_t}
\mathcal{S}(Y|\mu,\Sigma,v)=\frac{\Gamma(\frac{v+d}{2})|\Sigma|^{-0.5}}{(\pi
v)^{0.5d}\Gamma(\frac{v}{2})\{1+M_d(Y,\mu,\Sigma)/v\}^{0.5(v+d)}},
\end{equation}
where
\begin{equation}
M_d(Y,\mu,\Sigma)=(Y-\mu)^T \Sigma^{-1}(Y-\mu)
\end{equation}
denotes the Mahalanobis squared distance from $Y$ to $\mu$ with
respect to $\Sigma$.

As a mixture of Student's $t$ pdfs, $q$ is defined to be:
\begin{equation}\label{t-mixture}
q(x|\psi)\triangleq\sum\limits_{m=1}^M \alpha_m
\mathcal{S}(x|\mu_m,\Sigma_m,v),
\end{equation}
where $m$ denotes the component index, $M$ the total number of the mixing components, $\alpha_m$ the probability mass of the $m$th component satisfying $\alpha_{m}>0$ for each $m\in\{1,2,\ldots,M\}$ and $\sum_{m=1}^M\alpha_{m}=1$, and $\psi\triangleq\left\{M,\{\alpha_{m},\xi_m\}_{m=1}^M\right\}$ consists of all the tunable parameters of $q$.

The main task of the posterior exploration procedure, which will be presented in detail in the following subsection, is to tailor the shape of $q$ by updating the value of $\psi$, in order to make $q$ resemble the target distribution to the greatest extent.
\subsection{The posterior exploration procedure}\label{sec:PE}
The PE procedure is where our algorithm mainly differs with existing alternatives. The purpose of this procedure is to facilitate exploring target distribution structures that have not been represented enough by the IS pdf. As the target distribution is termed the posterior in Bayesian statistics, we call this procedure posterior exploration (PE).

The PE procedure runs four operators in turn, namely IS, componentwise Metropolis move \cite{haario2005componentwise}, expectation-maximization (EM) estimation and addition of new mixing components. Now we introduce the operators one by one in the following.
\subsubsection{The IS operator}\label{sec:IS}
IS is one of the basic Monte Carlo techniques for simulating a target distribution \cite{geweke1989bayesian,oh1992adaptive}. The input of this operator consists of a IS pdf $q(|\psi)$, a target pdf $\pi$ and the sample size $N$. Given the input, the IS operator first draws a set of random samples, $x_k^i$, $i=1,2,\ldots,N$, from $q$, and then weights the samples as follows,
\begin{equation}
\tilde{\omega}^i= \frac{\pi(x^i)}{q(x^i)}, i=1,2,\ldots,N.
\end{equation}
A normalization operation is finally performed,
\begin{equation}
\omega^i= \frac{\tilde{\omega}^i}{\sum_{i=1}^N\tilde{\omega}^i}, i=1,2,\ldots,N.
\end{equation}
The output of the IS operator is just a weighted sample set $\{x^i, \omega^i\}_{i=1}^N$.
\subsubsection{the componentwise Metropolis moving operator}
The input of this operator consists of the weighted sample set $\{x^i, \omega^i\}_{i=1}^N$ and the corresponding target pdf $\pi$.
The purpose of this operator is to explore the target distribution in a microscopic scale.
This operator gives a fine tuning to each sample.
Given the $i$th sample $x^i$, the operator tunes each dimension of it one by one. Suppose that the $j$th dimension is under consideration, the operator generates a zero-mean Gaussian distributed random number, and then adds it to the $j$th dimension of $x^i$ to generate a new sample. We denote the new sample by $x^{i\prime}$ and calculate the acceptance probability
\begin{displaymath}
\rho^i=\min\left\{\frac{\pi(x^{i\prime})}{\pi(x^i)},1\right\}.
\end{displaymath}
Finally we accept
\begin{displaymath}
x^i=\left\{\begin{array}{ll}
x^{i\prime}, \mbox{with probability}\quad \rho^i;\\
x^i, \mbox{with probability}\quad 1-\rho^i. \end{array} \right.
\end{displaymath}
\subsubsection{the EM operator}\label{sec:EM}
The EM mechanism is an effective approach to improve the IS efficiency by tailoring a mixture formed IS density function \cite{cappe2008adaptive}.
Inspired by the success of the application of the EM operator in the adaptive IS algorithm \cite{cappe2008adaptive}, we adopt it here to improve the IS efficiency of the PE-SMC algorithm. The input of the EM operator consists of a weighted sample set
$\{x^i,\omega^i\}_{i=1}^N$, the corresponding IS pdf $q(\cdot|\psi)$ and the target pdf $\pi$. The EM operator includes an Expectation and an Maximization step as follows \cite{cappe2008adaptive}:

the Expectation step:
\begin{equation}\label{eqn_alpha}
{\alpha_m}'=\sum\limits_{n=1}^N
w^i\epsilon(m|x^i).
\end{equation}

the Maximization step:
\begin{eqnarray}
{\mu_m}'&=&\frac{\sum\limits_{i=1}^N w^i
\epsilon(m|x^i)u_m(x^i)x^i }{\sum\limits_{i=1}^N \omega^i
\epsilon(m|x^i)u_m(x^i)},\\
{\Sigma_m}'&=&\frac{\sum\limits_{i=1}^N \omega^i
\epsilon(m|x^i)u_m(x^i)C_i }{\sum\limits_{i=1}^N \omega^i
\epsilon(m|x^i)},
\end{eqnarray}
where
\begin{equation}
u_m(x)=\frac{v+d}{v+(x-\mu_m)^T(\Sigma_m)^{-1}(x-\mu_m)},
\end{equation}
\begin{equation}\label{eqn:psi}
\epsilon(m|x)=\frac{\alpha_m \mathcal{S}(x|\mu_m,\Sigma_m,v)}{q(x|\psi)},
\end{equation}
and $C_i=(x^i-\mu_m)(x^i-\mu_m)^T$, $d$ denotes the dimension of $x$, and $A'$ denotes the updated value of $A$. Note that $\epsilon(m|x)$ in Eqn. (\ref{eqn:psi}) denotes the probability mass in the event that $x$ belongs to component $m$, and ${\alpha_m}'$ in Eqn. (\ref{eqn_alpha}) is just a weighted summation of those probability masses. Finally, update the parameters $\alpha_m={\alpha_m}', \mu_m={\mu_m}', \Sigma_m={\Sigma_m}'$, $m=1,\ldots,M$.
More details about the EM operation can be found for example in \cite{cappe2008adaptive,mclachlan2007algorithm,mclachlan2004finite}.
\subsubsection{Operator for adding new mixing components}
\begin{figure}[H]
\begin{tabular}{c}
\centerline{\includegraphics[width=3.5in,height=2.5in]{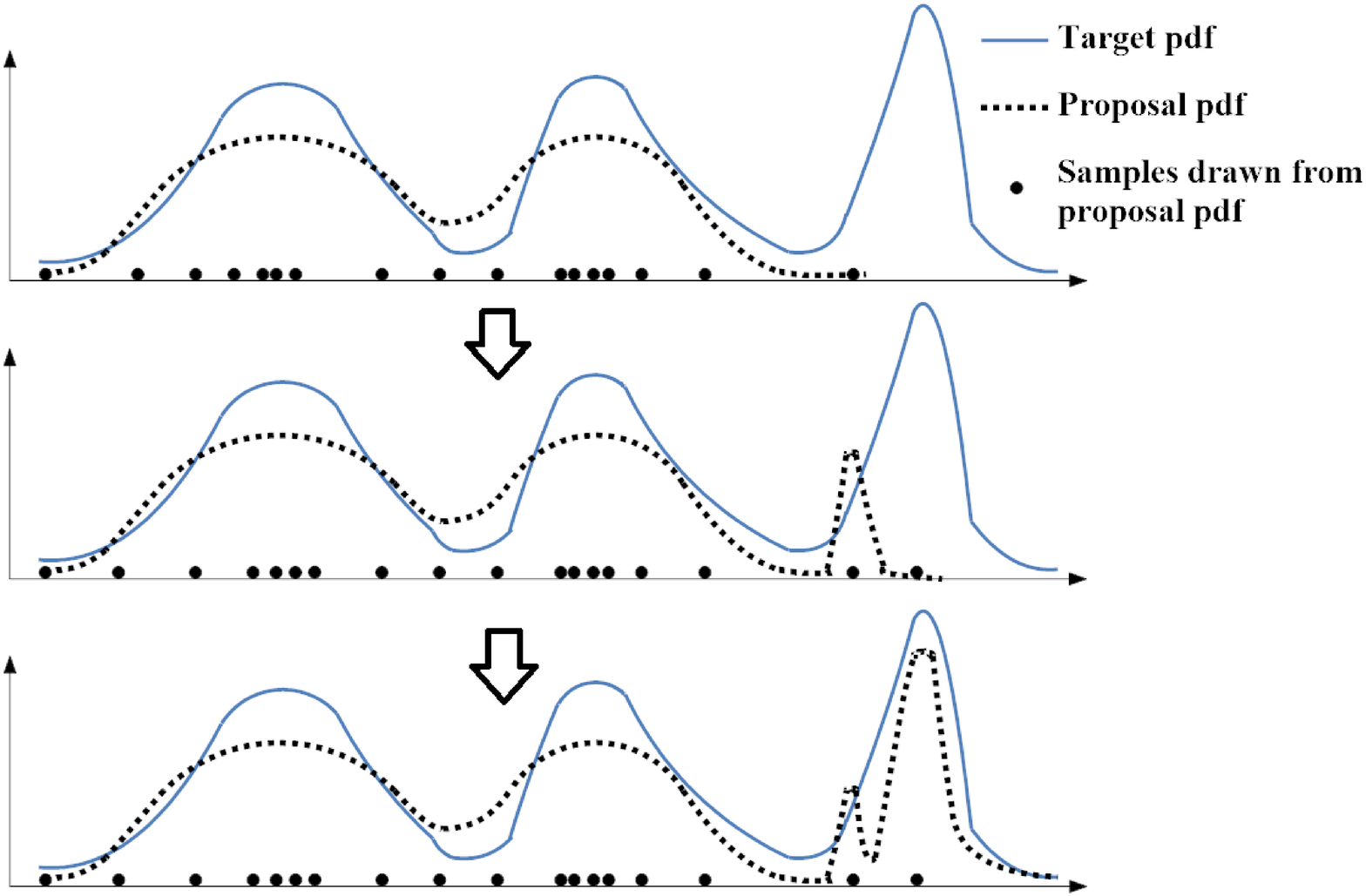}}
\end{tabular}
\caption{A conceptual illustration on the operator of adding new mixing components in the IS pdf}\label{fig:PE_concept}
\end{figure}

The purpose of this operator is to explore the promising solution space that has not been represented well enough or even totally neglected by the current IS pdf so far. This operator was originally developed by us in \cite{liu2014adaptive}.
The basic idea is illustrated conceptually in Fig. \ref{fig:PE_concept}. For simplicity and clarity in presenting the idea, a one dimensional case is considered.

The target and the IS pdfs are represented by the solid and dotted curves, respectively. The samples drawn from the IS pdf are represented by solid circles. As is shown in the top figure, at the beginning, the IS pdf does not represent the target structure well enough, since the rightmost peaky mode is not accounted. The above information is revealed by the highest weight sample. As seen in the top figure, the highest weight sample is the rightmost one. The idea is to add a new component centered at the highest weight sample into the mixture IS pdf and then draw a set of samples from the new component, as illustrated in the middle figure. The new samples that come from the new mixing component will further be used to explore the target. Again a new component is added and centered at the highest weight sample. Finally we obtain the bottom plot in Fig.\ref{fig:PE_concept}. It is shown that the operator of adding new mixing component can bring in a kind of ripple effect, which facilitates the algorithm to explore more promising areas in the solution space and finally an IS pdf that can represent the whole structure of the target will be obtained.

In practical implementation, after every ten new components being added, the total parameters of the mixture IS pdf are updated by running the IS and EM operators in turn, as described in Subsections \ref{sec:IS} and \ref{sec:EM}, respectively. The components that have negligible probability masses will be deleted and correspondingly probability masses of
the survival components are increased proportionally with respect to their
original values under the constraint that their summation is 1.

The stopping criteria of adding new components is the effective sample size (ESS), which is an empirical measure to evaluate the IS efficiency \cite{liu1998sequential}. Given a weighted sample set $\{x^i, \omega^i\}_{i=1}^N$, the ESS is defined to be
\begin{equation}
\mbox{ESS}\triangleq\frac{1}{\sum_{i=1}^N(\omega^i)^2}.
\end{equation}
and the normalized ESS (NESS) is ESS$/N$, which satisfies $1\leq\mbox{NESS}\leq1$. The limiting case $\mbox{NESS}=1$ happens only if the IS pdf is designed to be the same as the target and thus all the samples own the same weight $1/N$. So we use the NESS to measure the quality of the IS pdf based on the fact that the bigger the NESS the greater extent to which the IS pdf resembles the target. In practice, we check whether the NESS after adding each new mixing component is bigger than a prescribed threshold, if so, stop the process of adding new components, otherwise continue the process.
\subsection{Adaptive determination of the cooling temperature}\label{sec:temp_adapt}
Here we present an adaptive approach to determine $\lambda_{k+1}$ at the end of the $k$th iteration of the PE-SMC algorithm.
Observe that the annealed target pdf $\pi(\cdot)$ is a function of $\lambda$ as shown in Equation (\ref{anneal_fun}).
Given the IS pdf $q(\cdot|\psi_{k})$ and the random samples drawn from it, the $k+1$th target, $\pi_{k+1}(\cdot)$, actually determines the importance weights, which then determines an ESS value. Thus the ESS can be treated as a function of $\lambda$ as follows
\begin{equation}
\mbox{ESS}\triangleq h(\lambda).
\end{equation}
Here, based on $\lambda_{k}$, we specify $\lambda_{k+1}$ as follows
\begin{equation}
\lambda_{k+1}=\underset{\lambda\in(\lambda_{k},\infty)}{\arg\min}\|h(\lambda)-\beta\cdot\mbox{ESS}_{k}\|^2,
\end{equation}
where $\mbox{ESS}_{k}$ is calculated based on importance weights that are associated with the target pdf $\pi_{k}(\cdot)$ and the IS pdf $q(\cdot|\psi_{k})$, and the parameter $\beta\in(0,1)$ is used to control the extent to which $\pi_{k+1}(\cdot)$ differs from $\pi_{k}(\cdot)$. Empirically we set $\beta$ to 0.8 to guarantee a smooth transition from $\pi_{k}(\cdot)$ to $\pi_{k+1}(\cdot)$.
\section{Convergence analysis}
In this section, we make a detailed convergence analysis for the PE-SMC algorithm. To begin with, we show that under our assumptions on $\chi$ and $f$, the target distribution
converges weakly to the uniform distribution on the set of optimal solutions as the annealing temperature variable $\lambda$ in Eqn.(\ref{anneal_fun})
increases to infinity. In particular, if there is only one unique optimum solution, it converges weakly to
a degenerate distribution concentrated on that solution.

Set the \emph{target distribution} $r$ to be the absolutely continuous probability measure with density $\pi$, that is
\begin{equation}
r(A)=\int_A\pi(x)dx, \forall A\in\mathfrak{B}
\end{equation}
where $\mathfrak{B}$ denotes the Borel $\sigma$-field on $\chi$. For $\varepsilon>0$, define the $\varepsilon$-level set as follows
\begin{equation}
\chi_{\varepsilon}=\{x\in\chi: f(x)\geq f^{\ast}-\varepsilon\}
\end{equation}
where $f^{\ast}$ is, as before, the global maximum of $f$ on $\chi$. Then we have
\newtheorem{theorem}{PROPOSITION}
\begin{theorem}
For all $\varepsilon>0$, $\lim\limits_{\lambda\rightarrow\infty}r_{\lambda}(\chi_{\varepsilon})=1.$
\end{theorem}
\emph{Proof}. Fix $\varepsilon$. Then
\begin{eqnarray}
r_{\lambda}(\chi_{\varepsilon})&=&1-r_{\lambda}(\chi-\chi_{\varepsilon})=1-\frac{\int_{\chi-\chi_{\varepsilon}}f(x)^{\lambda}dx}{\int_{\chi}f(z)^{\lambda}dz}\\
&\geq&1-\frac{\int_{\chi-\chi_{\varepsilon}}f(x)^{\lambda}dx}{\int_{\chi_{\varepsilon/2}}f(z)^{\lambda}dz}\\
&\geq&1-\frac{(f^{\ast}-\varepsilon)^{\lambda}\phi(\chi-\chi_{\varepsilon})}{(f^{\ast}-\varepsilon/2)^{\lambda}\phi(\chi_{\varepsilon}/2)}\\
&=&1-\left(\frac{f^{\ast}-\varepsilon}{f^{\ast}-\varepsilon/2}\right)^{\lambda}\frac{\phi(\chi-\chi_{\varepsilon})}{\phi(\chi_{\varepsilon/2})}.
\end{eqnarray}
where $\phi$ denotes the Lebesgue measure on $\mathbb{R}^n$. (Note that our assumptions on $\chi$ and $f$ guarantee that $\phi(\chi_{\varepsilon/2})>0$.) Thus $\lim\limits_{\lambda\rightarrow\infty}r_{\lambda}(\chi_{\varepsilon})=1$.

Proposition 1 suggests that, as $\lambda$ tends to infinity, $r_{\lambda}$ converges weakly to a distribution that concentrates on the set
\begin{equation}
\chi_{\varepsilon}=\{x\in\chi: f(x)\geq f^{\ast}-\varepsilon\}. \nonumber
\end{equation}

Now fix $\lambda>0$ and hence the target pdf $\pi_{\lambda}$. Let's analyze convergence properties of the proposed four operators that constitute the PE procedure.
The IS principle suggests that the weighted samples yielded by the IS operator can give a Monte Carlo approximation to the current target pdf $\pi_{\lambda}$ \cite{geweke1989bayesian,oh1992adaptive,del2006sequential}. As its target pdf is also $\pi_{\lambda}$, the follow-up componentwise Metropolis moving operator will bring the distribution of the weighted samples closer to $\pi_{\lambda}$ and thus combat the approximation error introduced by the IS operator. The IS principle also tells us that the efficiency of the IS operator depends on the choice of the IS density function $q(\cdot|\varphi)$. The EM mechanism has proven to be an effective way to improve the quality of an mixture formed IS density function \cite{cappe2008adaptive}. The main shortcoming of EM lies in that it doesn't allow the number of mixing components $M$ to be changed, which indicates that, if the value of $M$ is not selected appropriately beforehand, the EM operator will not find the optimal parameter value for the IS density function. The last operator, namely the new components addition operator, remedies the defect of the EM operator by providing a mechanism to tailer $M$ online. As this operator increases the IS pdf around the local area centered at the current maximum weight sample, it always increases the ESS value to some extent. Since ESS is the most recognized metric to measure the IS efficiency in SMC methods \cite{doucet2009tutorial,del2006sequential,neal2001annealed}, the validity of the operation of adding new components is guaranteed.

To summarize, at each iteration of PE-SMC, the IS operator gives an initial Monte Carlo approximation to $\pi_{\lambda}$, then the follow-up operators contribute to decreasing the approximation error based on different mechanisms.

Since the random samples simulated with PE-SMC approximate $\pi_{\lambda}$ at each iteration and we have already proved that $\pi_{\lambda}$ converges weakly to the point mass at the global optimum $x^{\ast}$, we can conclude that the random samples simulated with PE-SMC will finally converge to $x^{\ast}$ as well.
\section{Performance evaluation}\label{sec:simulation}
\subsection{Performance comparison with other alternatives}
We applied the proposed algorithm, as well as the Trelea type vectorized PSO \cite{trelea2003particle}, the PFO algorithm \cite{zhou2014particle,chen2013population} and the SMC-SA algorithm \cite{zhou2013sequential}, to search the global maximum of over a dozen standard benchmark functions, termed TF1, TF2, $\ldots$, TF17 in what follows. See details about these functions in Sec. \ref{sec:appendix}. For some of the test functions, we considered their two dimensional (2D), 5D, 10D, and 20D implementations in order to investigate the abilities of the algorithms in handling higher dimensional cases. Following \cite{parsopoulos2007parameter,trelea2003particle}, the swarm size of PSO was set to 50 in all cases, and the swarm was allowed to perform a maximum number of 10000 iterations. For the other algorithms under consideration, the particle size was set to 500 for 2D cases, 2000 for 5D cases, 5000 for 10D cases and 50000 for 20D cases. The degree of freedom $v$ of all the Student's $t$ components is fixed to be 5. The termination condition is that in the last 10 iterations, no better solution is found.

Each algorithm is run 100 times independently for each optimization task and the convergence results are listed in Table I.
Boldface and italic in the table indicate the best and the second best result obtained, respectively.
A simple statistics to summarize the comparison results is presented in Table II. As is shown, for all 23 cases under consideration, PSO yielded the best result for 12 times, while it yielded the worst one for 3 times. The proposed PE-SMC algorithm did the best for 10 cases and it is more attractive than the others in that it never produced the worst performance.
\begin{table*}\centering\small
\begin{tabular}{llccccc}
\hline %
\multicolumn{2}{c}{Test Problems}& Goal:$f(\mbox{x}^{\star})$ &PSO&PF&SMC-SA&PE-SMC\\\hline
TF1 &2D& 30 &\textbf{30$\pm0$}  &29.4447$\pm0.53$ &29.9943$\pm6\times10^{-3}$  &\emph{29.9989}$\pm6.14\times10^{-4}$ \\\hline
TF2 &2D&1.56261 & \textbf{1.5626$\pm2.23\times10^{-15}$}  & 1.5625$\pm6.14\times10^{-5}$ &1.5626$\pm7.80\times10^{-6}$ &\emph{1.5626}$\pm5.01\times10^{-7}$ \\\hline
TF3 &2D&1 &\emph{0.9974}$\pm1.26\times10^{-2}$  &0.9836$\pm1.60\times10^{-2}$ &\textbf{0.9978$\pm8.60\times10^{-3}$}&0.9960$\pm5.50\times10^{-3}$ \\\hline
TF4 &2D&2459.6407 &2410.6$\pm75.08$  &2429.9$\pm26.6326$ &\textbf{2436.9$\pm24.1397$}&\emph{2431.4}$\pm30.2447$ \\\hline
TF5 &2D&1000 &\textbf{999.9979$\pm3.30\times10^{-3}$}  &999.8415$\pm0.16$ &999.9058$\pm8.16\times10^{-2}$&\emph{999.9895}$\pm8.3\times10^{-3}$ \\\hline
TF6 &2D&19.2085 &\textbf{19.2085}$\pm4.04\times10^{-14}$  &19.1981$\pm1.33\times10^{-2}$ &19.2084$\pm1.42\times10^{-4}$&\emph{19.2085$\pm6.11\times10^{-6}$} \\\hline
TF7 &2D&100 &\textbf{100$\pm0$}  &99.9996$\pm5.64\times10^{-4}$ &\emph{100}$\pm4.95\times10^{-5}$&\textbf{100$\pm0$} \\\hline
TF8 &2D&450 &\textbf{450$\pm0$}  &449.9759$\pm2.92\times10^{-2}$ &\emph{449.9990}$\pm1.20\times10^{-3}$&\textbf{450$\pm0$} \\\hline
TF9 &2D&200 &\textbf{200}$\pm0$  &199.9651$\pm3.60\times10^{-2}$ &199.9977$\pm3.00\times10^{-3}$ &\emph{199.9999}$\pm1.12\times10^{-6}$ \\\cline{2-7}
    &5D&200 &\textbf{200}$\pm0$ &197.0448$\pm0.62$&199.9854$\pm3.67\times10^{-2}$&\emph{199.9997}$\pm4.38\times10^{-4}$\\\cline{2-7}
    &10D&200 &\emph{199.8007}$\pm0.4690$ &193.9105$\pm34.27$&199.6702$\pm4.10\times10^{-2}$&\textbf{199.9487}$\pm6.48\times10^{-3}$\\\cline{2-7}
    &20D&200 &\emph{199.7810}$\pm0.5015$ &188.6582$\pm44.83$&196.3677$\pm5.33\times10^{-1}$&\textbf{199.9228}$\pm1.13\times10^{-2}$\\\hline
TF10 &2D&1&\textbf{1}$\pm0$&\emph{1}$\pm3.25\times10^{-5}$ &1$\pm6.55\times10^{-5}$ &1$\pm3.77\times10^{-5}$ \\\hline
TF11 &2D&1800&1758.5$\pm59.22$&1777.6$\pm32.09$ &\emph{1800}$\pm3.19\times10^{-4}$ &\textbf{1800}$\pm1.35\times10^{-4}$ \\\hline
TF12 &2D&486.7309&\textbf{486.7309}$\pm5.98\times10^{-13}$&485.599$\pm1.33$ &\emph{486.7309}$\pm9.91\times10^{-6}$ &486.7298$\pm9.50\times10^{-3}$   \\\hline
TF13 &2D&120&\textbf{120}$\pm0$&119.9994$\pm5.16\times10^{-4}$ &119.9999$\pm1.63\times10^{-4}$ &\emph{120}$\pm9.66\times10^{-6}$   \\\hline
TF14 &2D&1.8$\times10^5$&\emph{1.8}$\times10^5\pm2.32\times10^{-11}$&1.8$\times10^5\pm1.37\times10^{-2}$ &1.8$\times10^5\pm8.4\times10^{-3}$ &\textbf{1.8$\times10^5\pm0$ }  \\\hline
TF15&2D&$\approx$509\cite{yao1999evolutionary}&508.9427$\pm0.50$&508.9830$\pm0.11$ &\emph{509.0020}$\pm1.93\times10^{-10}$ &\textbf{509.0020}$\pm1.50\times10^{-11}$   \\\hline
TF16&2D&1&0.7900$\pm0.4094$ & \emph{0.9974}$\pm5.30\times10^{-3}$ &0.5000$\pm0.50$ &\textbf{1}$\pm1.21\times10^{-7}$   \\\hline
TF17 &2D&1.8013 &\textbf{1.8013}$\pm2.89\times10^{-15}$&1.7982$\pm3.10\times10^{-3}$&1.8008$\pm7.04\times10^{-4}$&\emph{1.8013}$\pm1.67\times10^{-8}$ \\\cline{2-7}
    &5D&4.687658 &\emph{4.6624}$\pm4.01\times10^{-2}$ & 4.2103$\pm0.1648$&4.6267$\pm0.1616$    &\textbf{4.6875}$\pm2.81\times10^{-4}$\\\cline{2-7}
    &10D&9.66015 &\emph{9.4562}$\pm0.16$&5.9357$\pm0.2852$&9.3155$\pm0.8074$&\textbf{9.6596}$\pm1.77\times10^{-2}$\\\hline
\end{tabular}
\caption{Convergence results yielded from 100 independent runs of each algorithm on each test problem}\label{Table:convergence values}
\end{table*}
\begin{table}[h!]
\begin {center}
\begin{tabular}{c||c|c|c|c }
\hline Algorithms & PSO & PFO & SMC-SA & PE-SMC\\
\hline $\sharp$Best & 12 & 0 & 2 & 10 \\
\hline $\sharp$2nd Best & 6 & 2 & 5 & 9 \\
\hline $\sharp$Worst & 3 & 17 & 2 & 0 \\
\hline
\end{tabular}
\end {center}
\caption{Statistics of the performance comparison results. $\sharp$ denotes the number of events, e.g., $\sharp$Best denotes the number of the event that the corresponding algorithm yielded the best result.}\label{Table:0dB}
\end{table}

Observe that PE-SMC beats significantly PFO and SMC-SA, which demonstrates the performance gain obtained due to the application of the proposed PE operation in the SMC framework.

After scrutiny on Table I, one can see that, for all 10D and 20D cases and all but one 5D cases, PE-SMC performs the best, which is an indication of that PE-SMC is more preferable in dealing with higher dimensional optimization problems.
\subsection{Examination of the PE procedure}
\begin{figure}[htb]
\begin{tabular}{c}
\centerline{\includegraphics[width=3.5in,height=2in]{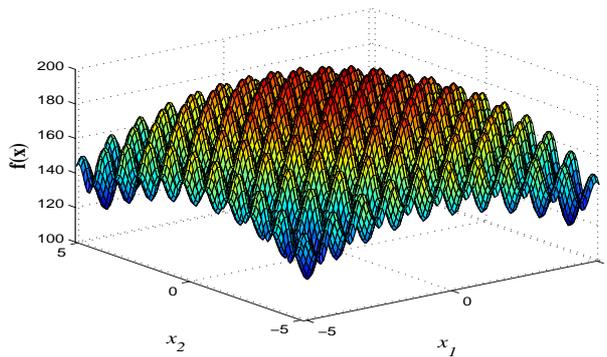}}
\end{tabular}
\caption{The 2D Rastrigin function defined in Sec.\ref{sec:rastrigin}}\label{fig:Rastrigin2D}
\end{figure}

For availability of visualization, we present intermediate results of applying the PE-SMC algorithm to search maximum of a 2D Rastrigin function defined in Sec.\ref{sec:rastrigin}, in order to examine the effect of the PE procedure. This test function is plotted in Fig.\ref{fig:Rastrigin2D}. The global maximum is 200 which is located at the origin point.

\begin{figure*}
\begin{tabular}{c}
\centerline{\includegraphics[width=5in,height=3in]{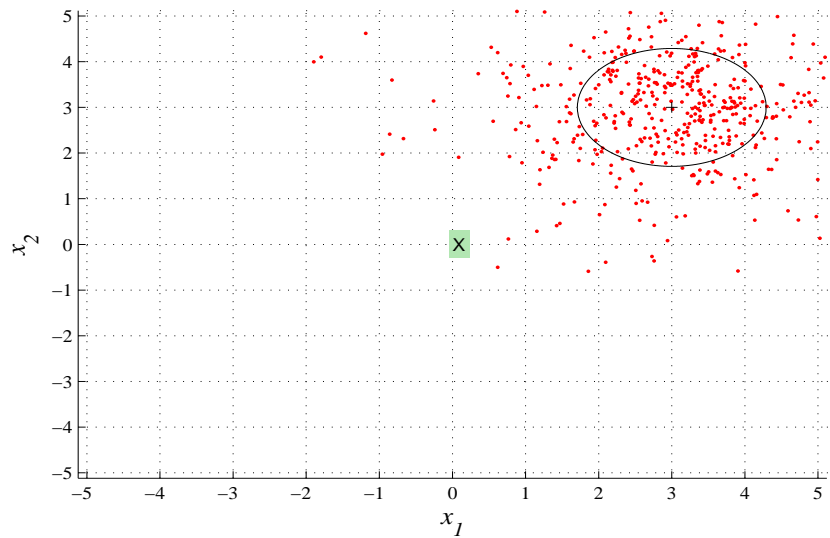}}
\end{tabular}
\caption{The IS pdf and the samples drawn from it in the 1st iteration before running the PE procedure}\label{fig:Rastr2D_init}
\end{figure*}

The PE-SMC algorithm is initialized by a one component IS pdf. The IS pdf as well as the samples drawn from it before running the PE procedure at the first iteration is plotted in Fig.\ref{fig:Rastr2D_init}. The contours in the figures represent the location, scale, and shape of the components in the Student's $t$-mixture IS pdf, which are drawn at one standard deviation from the means in each of the major and minor axis directions. As is shown, the current IS pdf does not cover well the important regions of the parameter space.

\begin{figure*}
\begin{tabular}{c}
\centerline{\includegraphics[width=5in,height=3in]{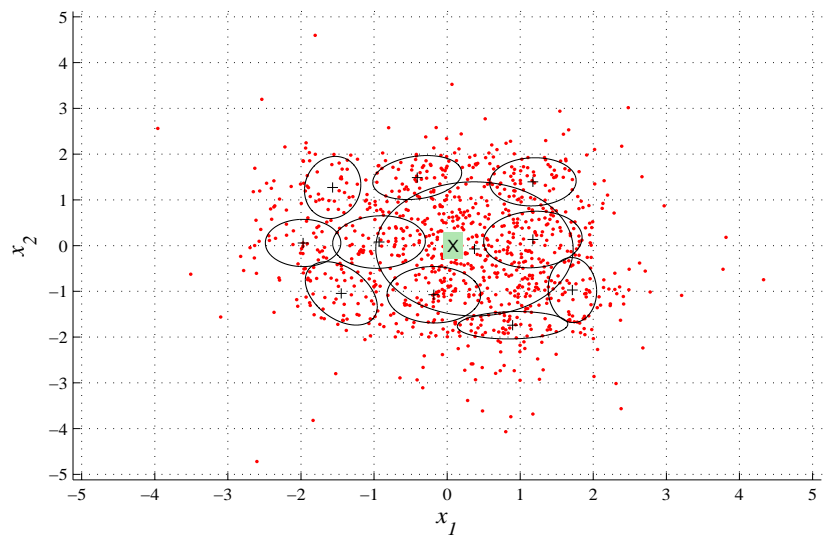}}
\end{tabular}
\caption{The IS pdf and samples drawn from it in the 1st iteration after running the PE procedure}\label{fig:Rastr2D_after_PE}
\end{figure*}

\begin{figure*}
\begin{tabular}{c}
\centerline{\includegraphics[width=5in,height=3in]{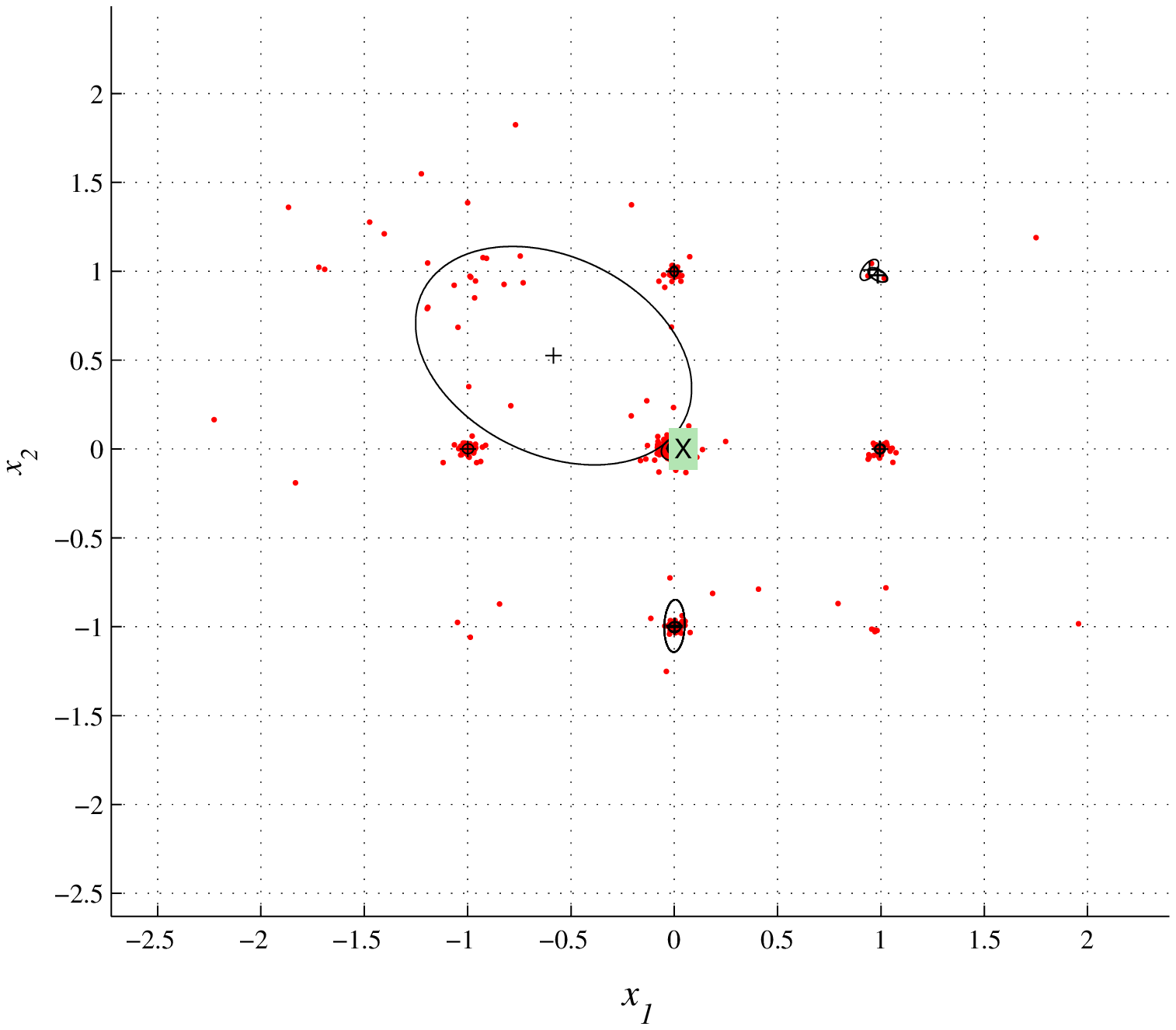}}
\end{tabular}
\caption{The IS pdf and samples drawn from it in the 4th iteration after running the PE procedure}\label{fig:Rastr2D_after_PE_4ter}
\end{figure*}

\begin{figure*}
\begin{tabular}{c}
\centerline{\includegraphics[width=5in,height=3in]{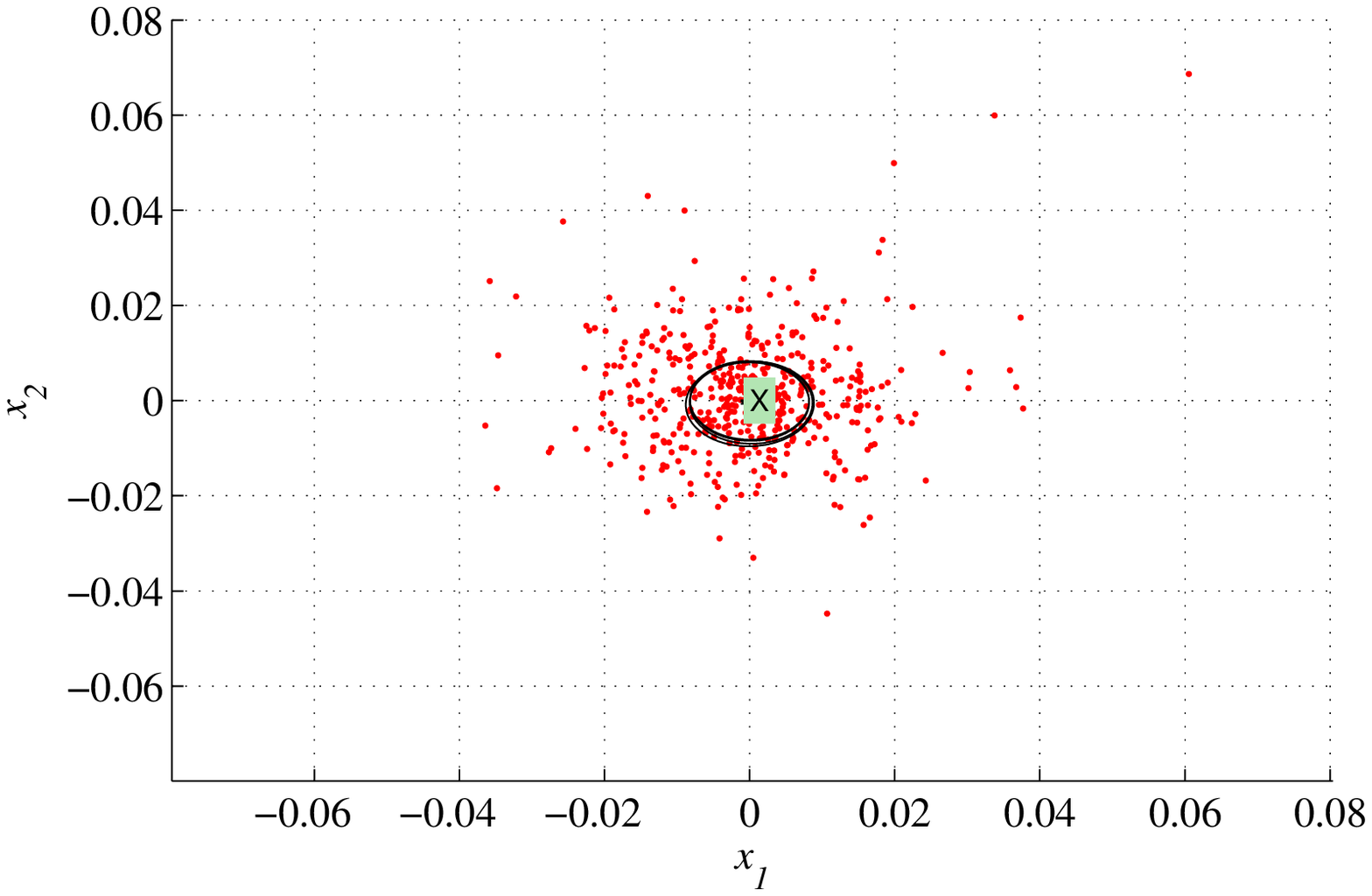}}
\end{tabular}
\caption{The IS pdf and samples drawn from it in the last iteration after running the PE procedure}\label{fig:Rastr2D_end}
\end{figure*}

The updated IS pdf and its corresponding samples after running the PE procedure are plotted in Fig.\ref{fig:Rastr2D_after_PE}. As is shown, the updated IS pdf owns more mixing components and its support has already covered the location of the global maximum. Because of the limited sample size and the multimodal structure of the test function as shown in Fig.\ref{fig:Rastrigin2D}, it is still nontrivial to find the global maximum. But as the algorithm iterates, the support of the IS pdf will converge on the local area of the maximum gradually, as shown in Figs.\ref{fig:Rastr2D_after_PE_4ter} and \ref{fig:Rastr2D_end}, which plot the results after running the PE procedure at the 4th and 7th iterations respectively. Note the scale changes of the figures.

\begin{figure*}
\begin{tabular}{c}
\centerline{\includegraphics[width=3.5in,height=2in]{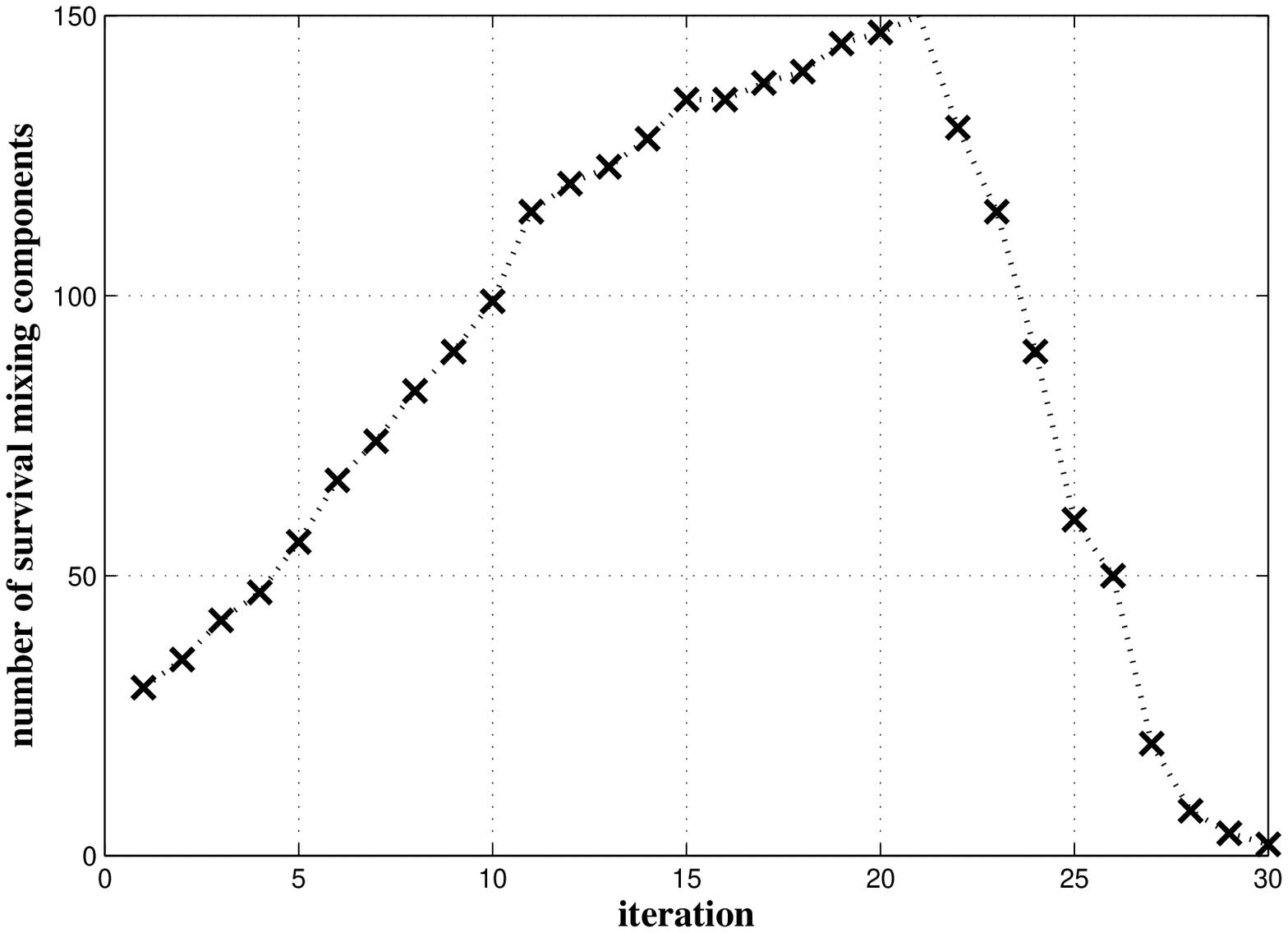}}
\end{tabular}
\caption{The number of survival mixing components in the IS pdf in an example run of the proposed PE-SMC algorithm in searching optimum for a 20D Rastrigin function defined in Sec.\ref{sec:rastrigin}}\label{fig:survival_vs_iter}
\end{figure*}

\begin{figure*}
\begin{tabular}{c}
\centerline{\includegraphics[width=3.5in,height=2in]{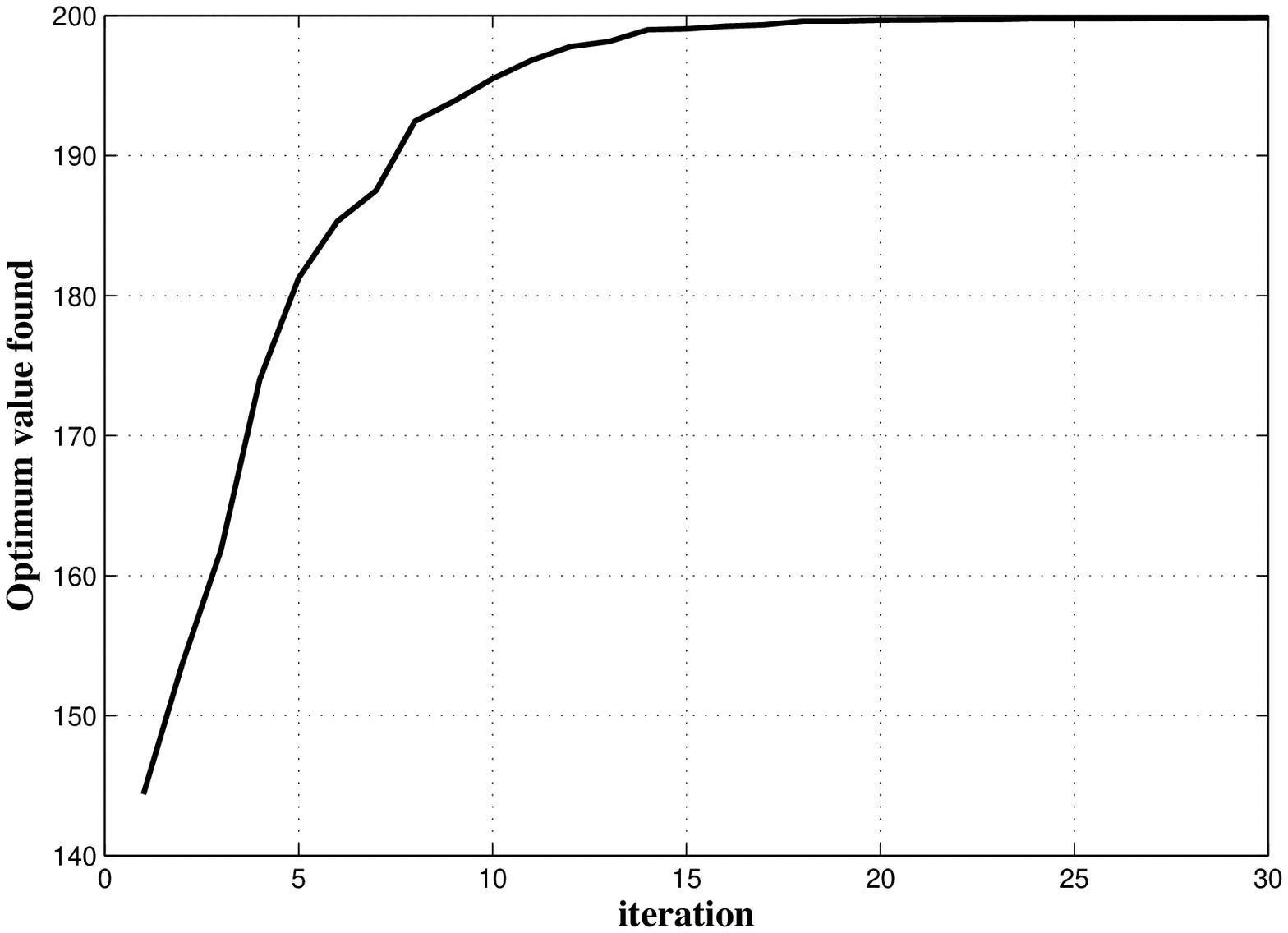}}
\end{tabular}
\caption{The value change of the optimum found as far in an example run of the proposed PE-SMC algorithm in searching optimum for a 20D Rastrigin function defined in Sec.\ref{sec:rastrigin}}\label{fig:opt_vs_iter}
\end{figure*}

For searching maximum of a 20D Rastrigin function, it is impossible to graphically plot the yielded intermediate IS pdf as before, while we recorded the number of survival mixing components in the IS pdf and now plot it in Fig.\ref{fig:survival_vs_iter}. The increase in the number of mixing components indicates an exploration dominated process, while the decrease corresponds to an exploitation dominated process. Taken together, it is shown that the proposed algorithm brings in a dynamic balance between the exploration and exploitation process. Thanks to this balance, the found optimum value of the test function converges fast to the goal value, 200, as shown in Fig.\ref{fig:opt_vs_iter}.
\subsection{Examination on the procedure of adapting the SA temperature schedule}
\begin{figure*}
\begin{tabular}{c}
\centerline{\includegraphics[width=3.5in,height=2in]{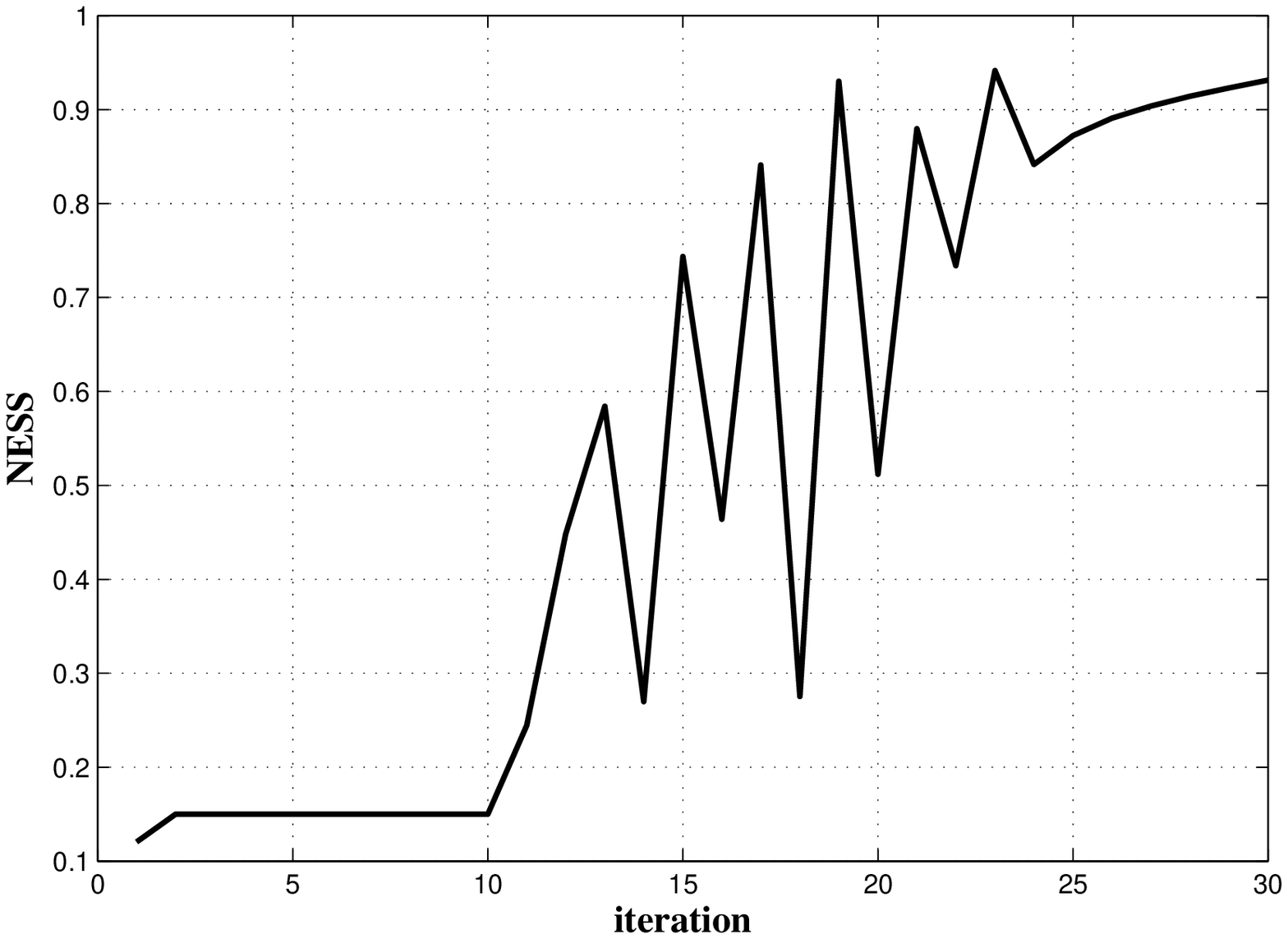}}
\end{tabular}
\caption{The normalized ESS (NESS) value in an example run of the PE-SMC algorithm with respect to the 20D Rastrigin test function defined in Sec.\ref{sec:rastrigin}}\label{fig:NESS_vs_iter}
\end{figure*}

\begin{figure*}
\begin{tabular}{c}
\centerline{\includegraphics[width=3.5in,height=2in]{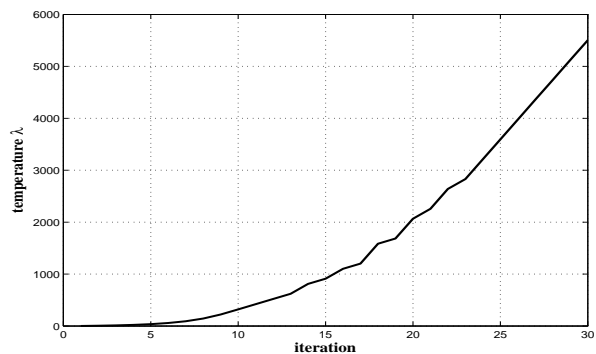}}
\end{tabular}
\caption{The yielded annealing temperature $\lambda$ of an example run of the PE-SMC algorithm with respect to the 20D Rastrigin test function}\label{fig:temp_vs_iter}
\end{figure*}

Take the result of an example run of our algorithm in a 20D Rastrigin function optimization task for example, the NESS changes per iteration as graphically shown in Fig.\ref{fig:NESS_vs_iter}. Observe that the NESS value fluctuates up and down, while the major trend is to become larger, which indicates that the IS efficiency grows along the adaptation of the SA's temperature. The yielded annealing temperature is plotted in Fig.\ref{fig:temp_vs_iter}. It is shown that the change of the annealing temperature almost conforms with an exponential growth law. We argue that this result is consistent with the theory which states that an exponentially long annealing schedule can guarantee convergence to the global optimum (even for non-convex problems) \cite{hajek1988cooling}.
\section{Concluding remarks}
A novel stochastic simulation based global optimization algorithm is proposed, which falls within the SMC framework. The key element of this method, which discriminates itself from the other existing alternatives, is an operator termed posterior exploration. This operator makes full use of the intermediate information yielded by the sequential importance sampling process, facilitating the exploration process to search potential promising regions in both macroscopic and microscopic scales. This algorithm also provides a practical solution to tradeoff the annealing temperature change rate and the sample size. Specifically, given a preset sample size, an adaptive method based on the IS efficiency measure ESS is proposed to determine the cooling schedule online. The algorithm was applied to over a dozen benchmark test functions and compared with existing alternatives in performance. The results show that our algorithm performs satisfactorily and consistently in all cases. In particular, it outperforms the PFO and SMC-SA methods significantly and is preferable to the PSO algorithm for higher dimensional cases.

Despite being convergent in theory, the PE-SMC algorithm still has limitations, which need further investigations in the future.
First, the sequence of annealed target pdfs converge on the set of global optima as the annealing variable $\lambda$ tends to infinity, while, in practice, the algorithm will end provided that it has no better solutions output for several successive iterations. Therefore, the sampling process usually stops at a relatively big $\lambda$ value (with respect to the initialized value for $\lambda$), other than infinity. How to set an appropriate initial $\lambda$ value also deserves further considerations. An empirical guideline is that, if the objective function $f(\cdot)$ is essentially plateaus
with a barely perceptible central peak, an initial $\lambda\gg1$ is preferable; otherwise, if $f(\cdot)$ is much peaky, then an initial $\lambda\ll1$ is better for use. Second, as an empirical measure to evaluate the IS efficiency, the ESS may become misleading under certain circumstances, e.g., when all the samples get trapped into a local region around a local optimum that is completely isolated from the true global optima.
Third, despite the great flexibility, the student's $t$ mixture pdf is not universally adaptable to all cases. For example, if the global optima happen to lie at the boundary of the solution space, asymmetric or truncated symmetric pdfs will be better choices for constructing the IS pdf in concept. Last but not least, the current PE-SMC algorithm still suffers from curse of dimensionality. How to adapt it to deal with ultra high dimensional cases with acceptable computational burdens is still an open problem.
%
\section{APPENDIX: TEST FUNCTIONS}\label{sec:appendix}
The original forms of these test functions are devised to find minimums. We translate them correspondingly, making them suitable for testing the algorithms that are devised to find the maximum. Assume that the original function $g(\mbox{x})$ defined on $\chi$ is upper bounded by $g_u$, i.e., $g(\mbox{x})\leq g_u, \forall \mbox{x}\in\chi$, then we just select a $g_s$ that is bigger than $g_u$ and translate the test function to be $f(\mbox{x})=g_s-g(\mbox{x})$. The modified test function is presented as follows.
\subsection{TF1: Ackley function}
This function is defined to be
\begin{equation}
f(\mbox{x})=30-\left[-a\exp\left(-b\sqrt{\frac{1}{d}\sum_{i=1}^dx_i^2}\right)-\exp\left(\frac{1}{d}\sum_{i=1}^d\cos(cx_i)\right)+a+\exp(1)\right].
\end{equation}
It is characterized by a nearly flat outer region, and a large peak at the centre, thus it poses a risk for optimization algorithms to be trapped in one of its many local maxima. The global maximum is $f(\mbox{x}^{\star})=30$, at $\mbox{x}^{\star}=(0,\ldots,0)$.
\subsection{TF2: Cross-in-tray function}
This function is defined to be
\begin{equation}
f(\mbox{x})=-0.5-\left(-0.0001\left(\left|g(x_1,x_2)\right|+1\right)^{0.1}\right),
\end{equation}
where
\begin{equation}
g(x_1,x_2)=\sin(x_1)\sin(x_2)\exp\left(\left|100-\frac{\sqrt{x_1^2+x_2^2}}{\pi}\right|\right). \nonumber
\end{equation}
It is evaluated on the square $x_i\in[-10,10]$, for $i = 1,2.$  The global maximum $f(\mbox{x}^{\star})=1.56261$ is located at $\mbox{x}^{\star}=$ $(1.3491,-1.3491),$ $(1.3491,1.3491),$ $(-1.3491,1.3491)$ and $(-1.3491,-1.3491)$.
\subsection{TF3: Drop-wave function}
This is multimodal and highly complex function. It is evaluated on the square $x_i\in[-5.12,5.12]$, for $i = 1,2$, as follows
\begin{equation}
f(\mbox{x})=\frac{1+\cos(12\sqrt{x_1^2+x_2^2})}{0.5(x_1^2+x_2^2)+2}.
\end{equation}
The global maximum $f(\mbox{x}^{\star})=1$ is located at $\mbox{x}^{\star}=(0,0)$.
\subsection{TF4: Eggholder function}
This is a difficult function to optimize, because of a large number of local minima. It is evaluated on the square $x_i\in[-512,512]$, for $i = 1,2,$ as follows
\begin{equation}
f(\mbox{x})=1500-\left[-(x_2+47)\sin\left(\sqrt{\left|x_2+\frac{x_1}{2}+47\right|}\right)-x_1\sin\left(\sqrt{|x_1-(x_2+47)|}\right)\right].
\end{equation}
The global maximum $f(\mbox{x}^{\star})=2459.6407$ is located at $\mbox{x}^{\star}=(512,404.2319)$.
\subsection{TF5: Griewank function}
This function has many widespread local minima, which are regularly distributed. It is usually evaluated on the hypercube $x_i\in[-600,600]$, for all $i = 1,\ldots,d$, as follows
\begin{equation}
f(\mbox{x})=1000-\left(\sum_{i=1}^d\frac{x_i^2}{4000}-\prod_{i=1}^d\cos\left(\frac{x_i}{\sqrt{i}}\right)+1\right).
\end{equation}
The global maximum $f(\mbox{x}^{\star})=1000$ is located at $\mbox{x}^{\star}=(0,\ldots,0)$.
\subsection{TF6: Holder table function}
This function has many local maximum, with four global one. It is evaluated on the square $x_i\in[-10,10]$, for $i = 1,2$, as follows
\begin{equation}
f(\mbox{x})=\left|\sin(x_1)\cos(x_2)\exp\left(\left|100-\frac{\sqrt{x_1^2+x_2^2}}{\pi}\right|\right)\right|.
\end{equation}
Its global maximum $f(\mbox{x}^{\star})=19.2085$ is located at $\mbox{x}^{\star}=(8.05502,9.66459),(8.05502,-9.66459),(-8.05502,9.66459)$ and $(-8.05502,-9.66459)$.
\subsection{TF7: Levy function}
This function is evaluated on the hypercube $x_i\in[-10,10]$, for all $i = 1,\ldots,d$, as follows
\begin{equation}
f(\mbox{x})=100-\left(\sin^2(\pi w_1)+\sum_{i=1}^{d-1}(w_i-1)^2[1+10\sin^2(\pi w_i+1)]+(w_d-1)^2[1+\sin^2(2\pi w_d)]\right).
\end{equation}
where $w_i=1+\frac{x_i-1}{4}$, for all $i=1,\ldots,d$.
The global maximum $f(\mbox{x}^{\star})=100$ is located at $\mbox{x}^{\star}=(1,\ldots,1)$.
\subsection{TF8: Levy function N.13}
This function is evaluated on the hypercube $x_i\in[-10,10]$, for $i = 1,2$, as below
\begin{equation}
f(\mbox{x})=450-\left(\sin^2(3\pi x_1)+(x_1-1)^2[1+\sin^2(3\pi x_2)]+(x_2-1)^2[1+\sin^2(2\pi x_2)]\right),
\end{equation}
with global maximum $f(\mbox{x}^{\star})=450$, at $\mbox{x}^{\star}=(1,\ldots,1)$.
\subsection{TF9: Rastrigin function}\label{sec:rastrigin}
This function is evaluated on the hypercube $x_i\in[-5.12,5.12]$, for all $i = 1,\ldots,d$, with the form
\begin{equation}
f(\mbox{x})=200-\left(10d+\sum_{i=1}^d[x_i^2-10\cos(2\pi x_i)]\right).
\end{equation}
Its global maximum $f(\mbox{x}^{\star})=200$ is located at $\mbox{x}^{\star}=(0,\ldots,0)$.
This function is highly multimodal, but locations of the local minima are regularly distributed.
\subsection{TF10: The second Schaffer function}
This function is usually evaluated on the square $x_i\in[-100,100]$, for all $i = 1,2$. It has a form as follows
\begin{equation}
f(\mbox{x})=1-\left(0.5+\frac{\sin^2(x_1^2-x_2^2)-0.5}{[1+0.001(x_1^2+x_2^2)]^2}\right).
\end{equation}
Its global maximum $f(\mbox{x}^{\star})=1$ is located at $\mbox{x}^{\star}=(0,0)$.
\subsection{TF11: Schwefel function}
This function is also complex, with many local minima. It is evaluated on the hypercube $x_i\in[-500,500]$, for all $i = 1,\ldots,d$, as follows
\begin{equation}
f(\mbox{x})=1800-\left(418.9829d-\sum_{i=1}^dx_i\sin(\sqrt{|x_i|})\right).
\end{equation}
Its global maximum $f(\mbox{x}^{\star})=1800$ is located at $\mbox{x}^{\star}=(420.9687,\ldots,420.9687)$.
\subsection{TF12: Shubert function}
This function has several local minima and many global minima. It is usually evaluated on the square $x_i\in[-10,10]$, for all $i = 1,2$.
\begin{equation}
f(\mbox{x})=300-\left(\sum_{i=1}^5i\cos((i+1)x_1+i)\right)\left(\sum_{i=1}^5i\cos((i+1)x_2+i)\right).
\end{equation}
Its global maximum is $f(\mbox{x}^{\star})=486.7309$.
\subsection{TF13: Perm Function 0,d,$\beta$}
This function is evaluated on the hypercube $x_i\in[-d,d]$, for all $i = 1,\ldots,d$, as follows
\begin{equation}
f(\mbox{x})=120-\left(\sum_{i=1}^d\left(\sum_{j=1}^d(j+\beta)\left(x_j^i-\frac{1}{j^i}\right)\right)^2\right).
\end{equation}
Its global maximum $f(\mbox{x}^{\star})=120$ is located at $\mbox{x}^{\star}=(1,\frac{1}{2},\ldots,\frac{1}{d})$.
\subsection{TF14: Rosenbrock function}
This function is also referred to as the Valley or Banana function. The global maximum lies in a narrow, parabolic spike.
However, even though this spike is easy to find, convergence to the maximum is difficult \cite{picheny2013benchmark}. It has the following form
\begin{equation}
f(\mbox{x})=1.8e5-\sum_{i=1}^{d-1}[100(x_{i+1}-x_i^2)^2+(x_i-1)^2],
\end{equation}
and is evaluated on the hypercube $x_i\in[-5,10]$, for all $i = 1,\ldots,d$.
The global maximum $f(\mbox{x}^{\star})=1.8e5$ is located at $\mbox{x}^{\star}=(1,\ldots,1)$.
\subsection{TF15: The fifth function of De Jong}
The fifth function of De Jong is multimodal, with very sharp drops on a mainly flat surface. The function is evaluated on the square $x_i\in[-65.536,65.536]$, for all $i=1,2$, as follows
\begin{equation}
f(\mbox{x})=510-\left(0.002+\sum_{i=1}^25\frac{1}{i+(x_1-a_{1i})^6+(x_2-a_{2i})^6}\right)^{-1},
\end{equation}
where
\begin{eqnarray}
\mathbf{a}&=&\left(\begin{array}{ccccccccc}
-32&-16&0& 16 & 32 & -32 & 9 & 16 & 32\\
-32&-32&-32 & -32 & -32 & -16 & 32 & 32 & 32
\end{array}\right).\nonumber
\end{eqnarray}
The global maximum $f(\mbox{x}^{\star})\approx509$ \cite{yao1999evolutionary}.
\subsection{TF16: Easom function}
This function has several local maximum. It is unimodal, and the global maximum has a small area relative to the search space.
It is usually evaluated on the square $x_i\in[-100,100]$, for all i = 1, 2, as follows
\begin{equation}
f(\mbox{x})=0-\cos(x_1)\cos(x_2)\exp\left(-(x_1-\pi)^2-(x_2-\pi)^2\right).
\end{equation}
The global maximum $f(\mbox{x}^{\star})=1$ is located at $\mbox{x}^{\star}=(\pi,\pi)$.
\subsection{TF17: Michalewicz function}
This function has $d!$ local minima, and it is multimodal. The parameter $m$ defines the steepness of the valleys and ridges; a larger $m$ leads to a more difficult search. The recommended value of $m$ is $m=10$.
It is usually evaluated on the hypercube $x_i\in[0,\pi]$, for all $i=1,\ldots,d$, as follows
\begin{equation}
f(\mbox{x})=-\sum_{i=1}^d\sin(x_i)\sin^{2m}\left(\frac{ix_i^2}{\pi}\right).
\end{equation}
The global maximum in 2D case is $f(\mbox{x}^{\star})=1.8013$ located at $\mbox{x}^{\star}=(2.20,1.57)$, in 5D case is $f(\mbox{x}^{\star})=4.687658$ and in 10D case is $f(\mbox{x}^{\star})=9.66015$.
\bibliographystyle{IEEEtran}
\bibliography{mybibfile}
%
\end{document}